\documentclass[]{spie}  

\usepackage{amsmath,amsfonts,amssymb}
\usepackage{graphicx}

\usepackage[colorlinks=true, allcolors=blue]{hyperref}
\usepackage{multirow}
\title{Ultra-low noise laser and optical frequency comb-based timing system for the Black Hole Explorer (BHEX) mission}

\author[1,2]{Hannah Tomio}
\author[2]{Guangning Yang}
\author[2]{Holly F. Leopardi}
\author[2]{Kenji Numata}
\author[2]{Anthony W. Yu}
\author[3]{Andrew Attar}
\author[2]{Xiaozhen Xu}
\author[2]{Wei Lu}
\author[2]{Cheryl Gramling}
\author[4]{T. K. Sridharan}
\author[2]{Peter Kurczynski}

\affil[1]{Massachusetts Institute of Technology, 77 Massachusetts Avenue, Cambridge, MA 02139, USA}
\affil[2]{NASA Goddard Space Flight Center, 8800 Greenbelt Rd, Greenbelt, MD 20771, USA}
\affil[3]{Vescent Photonics, 14998 W 6th Avenue Suite 700, Golden, CO 80401, USA}
\affil[4]{National Radio Astronomy Observatory, 520 Edgemont Road, Charlottesville, VA 22903, USA}

\authorinfo{Further author information: (Send correspondence to H.T.) \\ H.T.: \url{tomio@mit.edu})}

\begin{document} 
\maketitle

\begin{abstract}
In this effort, we demonstrate the performance of a highly stable time reference for the proposed Black Hole Explorer (BHEX) mission, a space-based extension to the Event Horizon Telescope (EHT) Very Long Baseline Interferometry (VLBI) project. This precision timing system is based on the use of a space-qualified, ultra-low noise laser developed as part of the Laser Interferometer Space Antenna (LISA) mission as the timing reference, and an optical frequency comb to transfer the stability of this laser to the microwave regime for instrumentation use. We describe the implementation of this system and experimental setup to characterize the stability performance. We present the results of this experiment that demonstrate the performance of this system meets requirements for the BHEX mission.
\end{abstract}


\section{Introduction}
\label{sec:intro} 
Very Long Baseline Interferometry (VLBI) enables high resolution imaging of radio sources by combining signals from multiple widely-separated radio telescopes. These signals are correlated between pairs of antennas with various separations, or baselines, which along with the observing wavelength dictate the angular resolution of the resulting image. This technique requires the array of radio telescopes to operate simultaneously, receiving and coherently recording the incident radiation field from the distant source. 

A highly stable time reference is necessary at each of the telescopes to maintain coherence between the independently recorded signals and a synchronization system between these references can reduce processing time during the data analysis stage. Instability in the time reference impacts the phase coherence, limiting the integration time and reducing the performance of the VLBI system. 

Ground-based arrays of telescopes, exemplified by the Event Horizon Telescope (EHT), are typically equipped with active hydrogen masers as the time reference. These and other atomic clocks can achieve frequency stabilities of a few parts in $10^{14}$  ($1 \times 10^{-14}$) or better over 10s of seconds of integration time necessary for VLBI. Synchronization can be performed using the Global Navigation Satellite System (GNSS). For the EHT, the Global Positioning System (GPS) is used to synchronize the array of radio telescopes within tens of nanoseconds\cite{eht_collab_2019}.

Extending beyond ground-based telescope arrays and into space can improve the angular and time resolution of observations by enabling longer interferometric baselines than are possible on the Earth and sampling a wide range of Fourier spatial frequencies throughout the orbital motion\cite{kurczynski_spie_2022}. However, active hydrogen masers and other atomic clocks can be challenging for space-based radio telescopes due to the size, weight, and power (SWaP) constraints of the space platform. Smaller, lighter, and less power-intensive time references, which meet the frequency stability requirements of VLBI and are capable of operating in the space environment, are highly desirable for future space VLBI missions like the Black Hole Explorer (BHEX). 

In this effort, we demonstrate the performance of one option for the time reference of the BHEX mission: the use of a space-qualified, ultra-low noise laser that is currently being developed as part of the Laser Interferometer Space Antenna (LISA) mission, and an optical frequency comb to transfer the stability of this laser to the microwave regime for instrumentation use. We describe the implementation of the microwave down-conversion, in which the LISA cavity-stabilized laser is locked to an optical frequency comb to divide the optical frequency down to 100 MHz. The fractional frequency stability of the 100 MHz signal is measured using a phase noise analyzer that is referenced to a separate laboratory ultra-stable laser system. We present the results of this experiment, which demonstrates that the performance of this system meets the BHEX requirements.

\section{Black Hole Explorer (BHEX) Mission and Requirements}
\label{sec:bhex} 
The BHEX mission is a proposed space VLBI extension to the EHT. It seeks to utilize the expanded interferometer baseline to advance fundamental understanding of black holes – in particular, to observe with sufficient spatial resolution a finely structured ring of light in the image of a black hole. Called the photon ring, this characteristic feature is formed by photons that orbit near the event horizon before escaping and reaching the observer. Detailed images of these features will provide new insights into the structure of black holes and enable new, high precision tests of general relativity. Further information on the scientific objectives of the BHEX mission can be found in \citenum{ BHEX_Johnson_2024, BHEX_Akiyama_2024, BHEX_Marrone_2024, BHEX_Peretz_2024, BHEX_Lupsasca_2024, BHEX_Galison_2024, BHEX_Kawashima_2024, BHEX_Issaoun_2024} and detailed descriptions of the development of the BHEX subsystem components are provided in \citenum{BHEX_Marrone_2024, BHEX_Peretz_2024, BHEX_Wang_2024, BHEX_Sridharan_2024, BHEX_Rana_2024, BHEX_Tong_2024, BHEX_Srinivasan_2024}.

Requirements for the time reference are determined from the scientific specifications of the mission: the frequencies of the observations (100 GHz through 300 GHz), the tolerance to coherence loss (of 5 to 10\%), and the coherent integration time (on the order of 10s of seconds, limited by atmospheric fluctuation). As these are still provisional for the proposed mission, the initial requirements we seek to meet are a fractional frequency stability (or Allan deviation) of $1-5 \times 10^{-14}$ over $\sim$10s of seconds of integration time. A higher stability, over a longer time interval, is of course desirable. This would facilitate compatibility with future higher frequency observations and enable high coherence, which will improve sensitivity and the spatial resolution of the resulting image\cite{sridharan_precision_2022}. 

This performance can be approached or achieved with a variety of technologies beyond active hydrogen masers and the optical frequency division technique described in this paper. Oven-controlled crystal oscillators (OCXOs) or ultra-stable oscillators (USO), while unable to match the stability performance of other methods, have demonstrated an Allan deviation of $< 10^{-13}$ at 10 seconds\cite{accubeat_2022}. These references are highly SWaP-efficient and have demonstrated heritage in space missions\cite{BHEX_Johnson_2024}. The ultimate choice of time reference for the BHEX mission will be driven by a combination of performance and minimizing SWaP. More details on the tradespace can be found in \citenum{BHEX_Peretz_2024}.  

\section{Optical Frequency Division-based Time Reference}
\label{sec:ofd} 
Optical frequency division, in which the stability of an ultra-stable laser is down-converted to the microwave regime using an optical frequency comb, can generate extremely stable microwave signals (fractional frequency stability of $\sim 10^{-16}$) to serve as time and frequency references\cite{xie_2017}. 

In this approach, the optical frequency comb, a self-referenced femtosecond laser, is phase-locked to an ultra-low noise, cavity-stabilized continuous wave laser in order to stabilize the repetition rate frequency ($f_{rep}$) of the comb. This transfers the spectral purity and stability of the ultra-stable laser to the modes of the frequency comb. A fast photodetector is used to detect the stabilized pulse train and generate the microwave/radio frequency (RF) $f_{rep}$ and its harmonics. With filtering the desired microwave frequency, which maintains the fractional frequency stability of the optical reference, can be selected and then utilized for instrumentation. 

The technique of optical frequency division was first demonstrated in the mid-2000s, following the advent of optical frequency combs generated by femtosecond lasers\cite{bartels_2005, fortier_2011}. While revolutionary for time and frequency metrology, this approach has yet to be widely adopted in applications outside of a laboratory environment, as both the ultra-stable reference laser and optical frequency comb are typically SWaP-inefficient and relatively fragile. In this effort, we leverage recent developments in robust, low-SWaP, commercially available optical frequency combs and a concurrent effort to develop a space-qualified, ultra-low noise laser for the Laser Interferometer Space Antenna (LISA) mission. 

The LISA mission, led by ESA with contributions from NASA and ESA member states, is a space-based laser interferometry mission intended to measure gravitational waves, with a planned launch date in the mid-2030s. Consisting of three identical spacecraft, each separated by 2.5 million km to form an equilateral triangle, LISA will employ two laser links between each spacecraft to form heterodyne interferometers\cite{amaro-seoane_2017}. These interferometers must be sensitive to picometer-level variations in the optical path (of $\sim 2.5$ million km) at timescales of 1000 seconds. Hence, the LISA lasers must demonstrate extremely low levels of frequency and intensity noise for a space-based laser system.

NASA Goddard Space Flight Center (GSFC) is currently developing a laser transmitter intended for the LISA mission. The transmitter is based on a master oscillator power amplifier (MOPA) architecture, with a miniaturized non-planar ring oscillator (NPRO) as the master oscillator and a forward-pumped fiber amplifier as the power amplifier. A fiber-coupled electro-optic modulator (EOM) is inserted between the low-noise seed laser and amplifier to provide phase modulation of the sideband to transmit reference clock information. A packaged laser optical module with a technology readiness level (TRL) of 4 has already been completed, evaluated, and shown to meet the initial LISA requirements. Present efforts focus on the development of a TRL 6 module, more details of which can be found in \citenum{yu_lisa_2022, numata_lisa_2022}.  The LISA frequency reference system, based on a high finesse optical reference cavity, is also in development\cite{numata_lisa_2022}. 

\section{Experiment and Results}
\label{sec:experiment}
\begin{figure} [ht]
   \begin{center}
   \begin{tabular}{c} 
   \includegraphics[keepaspectratio,width=0.9\textwidth]{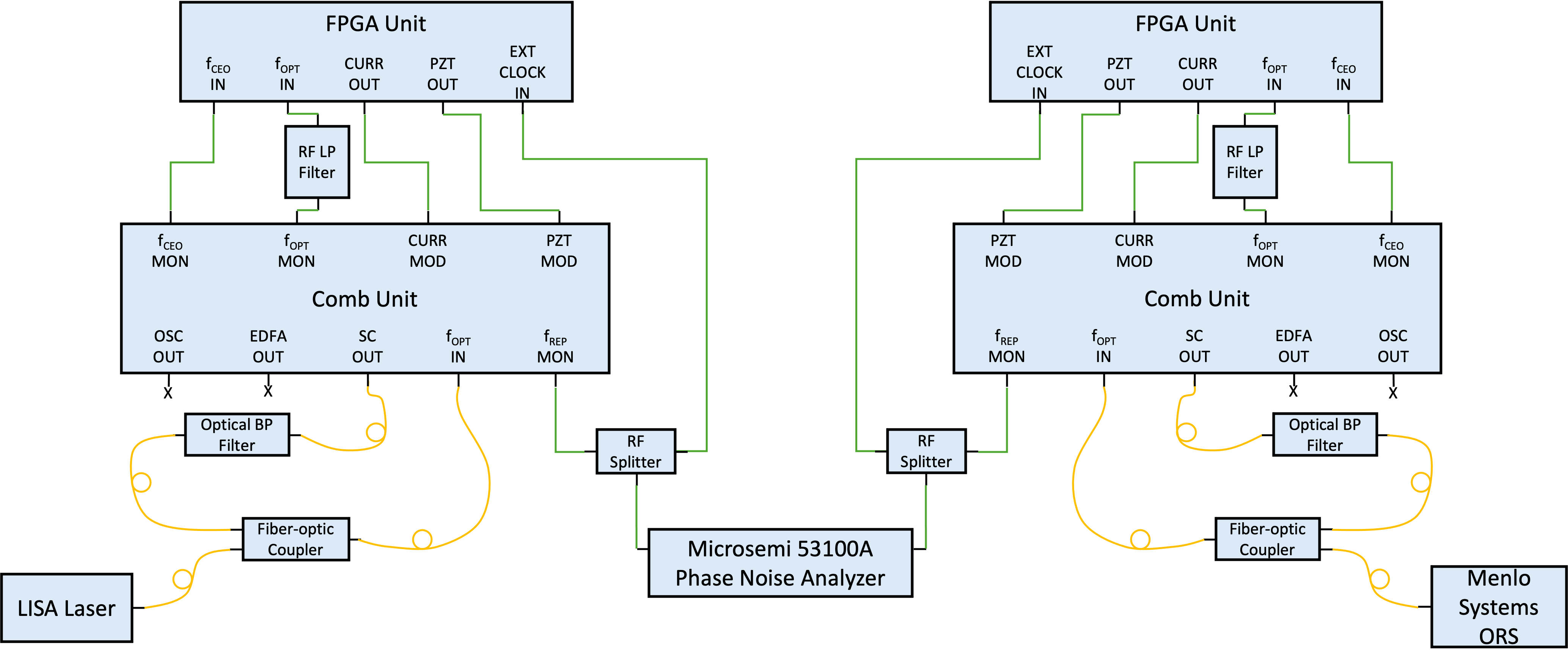}
   \end{tabular}
   \end{center}
   \caption{\label{fig:setup} 
Experimental setup with LISA laser, Menlo Systems ORS, and two optical frequency combs (Vescent FFC-100s) for optical frequency division
}
\end{figure} 

\begin{figure} [ht]
   \begin{center}
   \begin{tabular}{c} 
   \includegraphics[keepaspectratio,width=0.9\textwidth]{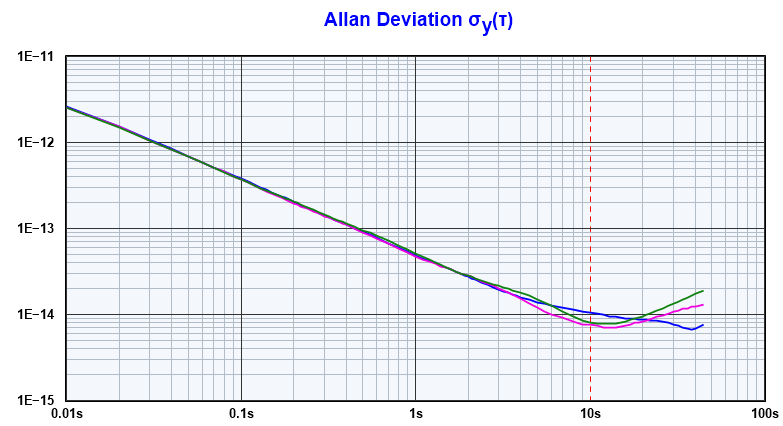}
   \end{tabular}
   \end{center}
   \caption{\label{fig:allan_dev} 
Allan deviation of the 100 MHz output generated via optical frequency division with LISA laser prototype
}
\end{figure} 
In this effort, we used a fiber frequency comb from Vescent Photonics (the FFC-100) to phase-lock to a prototype of the LISA laser transmitter in development at NASA GSFC. This prototype is stabilized to an optical reference cavity as part of the LISA ground support equipment (GSE). To generate a beatnote, the supercontinuum output of the FFC-100 is filtered with a narrow 0.5~nm bandpass filter and overlapped with the stabilized 1064.5~nm optical output from the LISA laser in a fiber-optic coupler. The beatnote signal ($f_{opt}$) is input to the FFC-100 for internal photodetection. The integrated FFC-100 contains optical and electrical signal conditioning necessary to generate a high signal-to-noise ratio RF $f_{opt}$ signal. It also generates and conditions the carrier-envelope offset (CEO) frequency ($f_{CEO}$) RF signal, which is describes the changing carrier envelope phase of the mode-locked laser. The feedback control loops for $f_{opt}$ and $f_{CEO}$ are handled by a separate field programmable gate array (FPGA)-based control unit (the SLICE-FPGA-II), which performs the digital proportional-integral-derivative (PID) lock on these signals to internal RF oscillators. The control signals (for current and high-speed PZT actuator) are input back to the FFC-100 to stabilize $f_{CEO}$ and $f_{opt}$, and thereby stabilize the mode-locked laser’s repetition rate, which is output as the 100 MHz $f_{rep}$ signal. 

As the phase noise of this signal is lower than that of most commercially available microwave references, a second, near-identical optical frequency division setup was built to characterize the stability of the generated $f_{rep}$ signal. This test setup featured the same model of FFC-100 fiber frequency comb locked to a Menlo Systems Optical Reference System (ORS). The relative stabilities of the 100 MHz signals were compared using a Microsemi 53100A Phase Noise Analyzer. The full experiment setup is shown in Figure~\ref{fig:setup}. Allan deviations of $1-5 \times 10^{-14}$ or less at the integration times ($\tau$) of interest (1, 5, 10, and 30 seconds) were achieved. These results from three runs of the experiment are shown in Figure~\ref{fig:allan_dev} and summarized in Table~\ref{tab:allan_dev}.

\begin{table}[ht]
\centering
\caption{\label{tab:allan_dev} 
Allan deviations at various averaging times ($\tau$) from Figure~\ref{fig:allan_dev}
}
\begin{tabular}{|l|lll|}
\hline
\multicolumn{1}{|c|}{\multirow{2}{*}{$\tau$}} & \multicolumn{3}{c|}{Allan Deviation}                            \\ \cline{2-4} 
\multicolumn{1}{|c|}{}                        & \multicolumn{1}{l|}{Run 1} & \multicolumn{1}{l|}{Run 2} & Run 3 \\ \hline
1 s                                            & \multicolumn{1}{l|}{$4.92 \times 10^{-14}$}     & \multicolumn{1}{l|}{$4.78 \times 10^{-14}$}     & $5.16 \times 10^{-14}$     \\ \hline
5 s                                            & \multicolumn{1}{l|}{$1.39 \times 10^{-14}$}     & \multicolumn{1}{l|}{$1.19 \times 10^{-14}$}     & $1.49 \times 10^{-14}$    \\ \hline
10 s                                           & \multicolumn{1}{l|}{$1.04 \times 10^{-14}$}     & \multicolumn{1}{l|}{$7.58 \times 10^{-14}$}     & $7.97 \times 10^{-15}$    \\ \hline
30 s                                           & \multicolumn{1}{l|}{$7.71 \times 10^{-15}$}     & \multicolumn{1}{l|}{$1.08 \times 10^{-14}$}     & $1.35 \times 10^{-14}$    \\ \hline
\end{tabular}
\end{table}

\section{Conclusion and Future Work}
\label{sec:conclusion}
We demonstrate the performance of one option for the time/frequency reference of a future space VLBI concept, the BHEX mission. The technique of optical frequency division is employed to transfer the stability of the ultra-stable LISA laser to the microwave regime using an optical frequency comb. While the results of this demonstration show this technique meets the preliminary BHEX requirements of a fractional frequency instability of $1-5 \times 10^{-14}$ over $\sim$10s of seconds of integration time, better performance over shorter averaging times can be expected with improvements to the test setup and measurement approach. Future work will pursue these improvements and seek to qualify the optical frequency comb for operation in the space environment. 

\acknowledgments 
This work was supported by a NASA Space Technology Graduate Research Opportunity (NSTGRO) and by the Internal Research and Development (IRAD) program at NASA Goddard Space
Flight Center.

\bibliography{references, bhexspiepapers} 
\bibliographystyle{spiebib} 

\end{document}